\documentclass[aps,prb,twocolumn,showpacs,superscriptaddress]{revtex4}
\usepackage{amssymb}
\usepackage{txfonts}
\usepackage{graphicx}

\begin{document}

\title{Electron-doping dependence of the anisotropic superconductivity in BaFe$_{2-x}$Ni$_{x}$As$_2$}

\author{Zhaosheng Wang}
\email{z.wang@hzdr.de}
\affiliation{Hochfeld-Magnetlabor Dresden (HLD-EMFL), Helmholtz-Zentrum Dresden-Rossendorf, D-01314 Dresden, Germany}

\author{Tao Xie}
\affiliation{Beijing National Laboratory for Condensed Matter
Physics, Institute of Physics, Chinese Academy of Sciences, Beijing
100190, China}

\author{E. Kampert}
\affiliation{Hochfeld-Magnetlabor Dresden (HLD-EMFL), Helmholtz-Zentrum Dresden-Rossendorf, D-01314 Dresden, Germany}

\author{T. F\"{o}rster}
\affiliation{Hochfeld-Magnetlabor Dresden (HLD-EMFL), Helmholtz-Zentrum Dresden-Rossendorf, D-01314 Dresden, Germany}

\author{Xingye Lu}
\affiliation{Beijing National Laboratory for Condensed Matter
Physics, Institute of Physics, Chinese Academy of Sciences, Beijing
100190, China}

\author{Rui Zhang}
\affiliation{Beijing National Laboratory for Condensed Matter
Physics, Institute of Physics, Chinese Academy of Sciences, Beijing
100190, China}

\author{Dongliang Gong}
\affiliation{Beijing National Laboratory for Condensed Matter
Physics, Institute of Physics, Chinese Academy of Sciences, Beijing
100190, China}

\author{Shiliang Li}
\affiliation{Beijing National Laboratory for Condensed Matter
Physics, Institute of Physics, Chinese Academy of Sciences, Beijing
100190, China}
\affiliation{Collaborative Innovation Center of Quantum Matter, Beijing, China}

\author{T. Herrmannsd\"{o}rfer}
\affiliation{Hochfeld-Magnetlabor Dresden (HLD-EMFL), Helmholtz-Zentrum Dresden-Rossendorf, D-01314 Dresden, Germany}

\author{J. Wosnitza}
\affiliation{Hochfeld-Magnetlabor Dresden (HLD-EMFL), Helmholtz-Zentrum Dresden-Rossendorf, D-01314 Dresden, Germany}

\author{Huiqian Luo}
\email{hqluo@iphy.ac.cn}
\affiliation{Beijing National Laboratory for Condensed Matter
Physics, Institute of Physics, Chinese Academy of Sciences, Beijing
100190, China}

\begin{abstract}
The upper critical field ($H_{c2}$) in superconducting
BaFe$_{2-x}$Ni$_{x}$As$_2$ single crystals has been determined by
magnetotransport measurements down to 0.6 K over the whole
superconducting dome with $0.065 \leqslant x \leqslant 0.22$, both
for the inter-plane ($H \parallel c$, $H_{c2}^{c}$) and in-plane ($H
\parallel ab$, $H_{c2}^{ab}$) field directions in static magnetic fields
up to 16 T and pulsed magnetic fields up to 60 T.  The temperature
dependence of $H_{c2}^{ab}$ follows the
Werthamer-Helfand-Hohenberg (WHH) model incorporating orbital and
spin paramagnetic effects, while $H_{c2}^{c}(T)$ can only be
described by the effective two-band model with unbalanced
diffusivity. The anisotropy of the upper critical fields,
$\gamma (T)=H_{c2}^{ab}/H_{c2}^{c}$ monotonically increases with
increasing temperature for all dopings, and its zero-temperature limit,
$\gamma (0)$, has an asymmetric doping dependence with a significant enhancement in the overdoped regime,
where the optimally doped compound has the most isotropic superconductivity.
Our results suggest that the anisotropy in the superconductivity of iron pnictides is
determined by the topology of the Fermi surfaces together with the doping-induced impurity scattering.

\end{abstract}


\maketitle

\section{Introduction}

Determining the upper critical field ($H_{c2}$), where superconductivity ceases in a type-II superconductor, is one of the
most important steps for gathering an understanding of unconventional
superconductivity including the pairing mechanism, the pairing
strength, as well as the coherence length.  Particularly, the
temperature dependence of $H_{c2}$ reflects the underlying
electronic structure responsible for superconductivity and provides
valuable information on the microscopic origin of pair breaking,
which is important for various application purposes, too. Most iron-based
superconductors have a moderate $H_{c2}$, within the range of
non-destructive pulsed high magnetic fields available in current technology
\cite{Hosono1,WangZSPRB,Jaroszynski,LeeHS,HunteF, Altarawneh,
YuanHQ, KanoM, Khim, Khim2}. This makes it possible to obtain
$H_{c2}$ and its anisotropy $\gamma=H_{c2}^{ab}/H_{c2}^{c}$ down to
the zero-temperature limit experimentally, rather than by use of imprecise theoretical extrapolations
from data near the superconducting transition temperature
($T_c$) only. Although all iron-based superconductors have a layered structure. The
122 and 11 families always show a nearly isotropic $H_{c2}(0)$ at zero temperature
[$\gamma(0) \approx 1$] \cite{Altarawneh, YuanHQ, KanoM, Khim}, and
the 111 family has a slightly higher $\gamma(0) \approx 1.5$
\cite{Khim2,YuanHQ2}, while the anisotropy of $H_{c2}(0)$ for the 1111
family with the highest $T_c$ is not well determined yet due to the very
high $H_{c2}$ (above 100 T)\cite{Jaroszynski,LeeHS,HunteF}. Moreover, it should be
noticed that, until now most results are from optimally doped compounds or
intrinsically superconducting samples and reports on the doping dependence of $H_{c2}$ and $\gamma$ at
low temperature in iron pnictides are scarce.

BaFe$_{2}$As$_2$ (Ba-122), as a parent phase of the iron-based
superconductors with a double-layered structure, can
be doped either with holes by replacing Ba with K/Na or with electrons by
substituting Fe with Ni/Co to suppress antiferromagnetism (AFM) and
induce superconductivity
\cite{RotterM1,RotterM2,XuZA,NiN,MandrusD,AswarthamS}. Specifically
in the BaFe$_{2-x}$Ni$_{x}$As$_2$ system, upon doping electrons by Ni
substitution, the N$\rm \acute{e}$el temperature ($T_N$) and
orthorhombic lattice distortion temperature ($T_s$) are
gradually suppressed and vanish at about $x=$ 0.11. Superconductivity emerges at
$x=$ 0.05, then reaches the maximum critical temperature ($T_c$) at about 20 K around $x=$ 0.1, and
finally disappears in the overdoped regime at $x=$ 0.25 [Fig.
1(a)] \cite{LuoHQ1,LuXY}. Based on a rigid-band model, the doped
electrons simply shift the chemical potential, which shrinks the
hole pocket at the $\Gamma$ point and enlarges the electron pocket at the M
point, respectively [Fig. 1(b)] \cite{DingH}. Although the real case
is more complicated due to additional effects from impurities
\cite{KuW,BlackburnS}, the experimental results qualitatively agree with the
rigid-band model \cite{IdetaS}. Moreover, all Fermi-surface sheets show
warping along the $k_z$ direction, and the hole sheet becomes more
three-dimensional (3D) upon electron doping into the overdoped
regime \cite{LiuC}. Thus, a more
isotropic superconductivity corresponding to the 3D
topology of the Fermi surfaces is generally expected in the electron-overdoped compounds. It is noticed that the hole-doped
Ba$_{1-x}$K$_{x}$Fe$_2$As$_2$ system shows a clearly increasing
anisotropy of $H_{c2}$ in overdoped samples \cite{LiuY} with a sign change of the superconducting order parameter across different Fermi-surface pockets\cite{ZhangS}, where a
Lifshitz transition occurs changing the Fermi-surface topology \cite{KhanSN}.
While these results are inspiring, electric-transport
measurements were only performed in static fields up to 9 T and,
thus, are unlikely to allow a conclusive determination on the nature
of $H_{c2}$ and $\gamma$ throughout the phase diagram.

\begin{figure}
\includegraphics[scale=0.27]{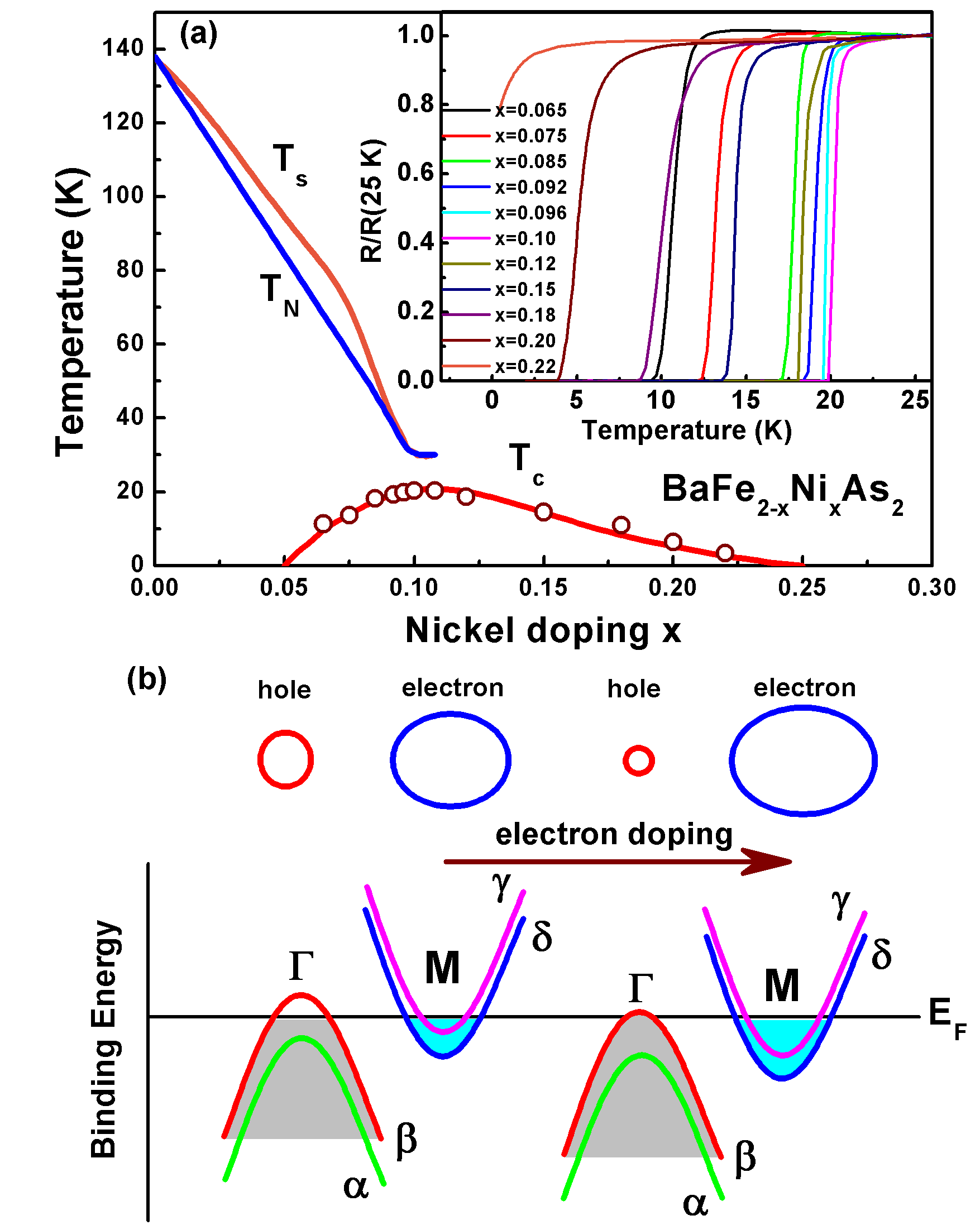}
\caption{(color online) (a) Phase diagram of
BaFe$_{2-x}$Ni$_{x}$As$_2$, where $T_s$, $T_N$, and $T_c$ are the
structural, AFM, and superconducting transition temperatures,
respectively. The open symbols mark the 12 doping levels studied in
this work.  The inset shows the sharp superconducting transition of
the in-plane resistance normalized at 25 K. (b) Schematic picture of
the electron-doping effects on the Fermi pockets and band structure
based on a rigid-band model. } \label{fig:fig1}
\end{figure}

Here, we report a systematic study of $H_{c2}$ and its anisotropy
in electron-doped BaFe$_{2-x}$Ni$_{x}$As$_2$ single crystals at
high magnetic fields up to 60 T and low temperatures down to 0.6 K.
We establish the doping and temperature dependence both for
$H_{c2}$ and $\gamma$ throughout the superconducting dome with
$0.065 \leqslant x \leqslant 0.22$. While our data for optimally doped samples are consistent with earlier results \cite{Altarawneh, YuanHQ, KanoM, Khim},
showing a nearly isotropic superconductivity with $\gamma (0)= 1.02$ for the zero-temperature limit,
we find that the anisotropy has a very asymmetric doping dependence and increases beyond 2 in the overdoped regime. Further
analysis of $H_{c2}(T)$ suggests that the
anisotropy is intimately related to the
change of the Fermi-surface topology together with
impurity scattering from the Ni dopants.

\section{Experiment}

\begin{figure}
\includegraphics[scale=0.55]{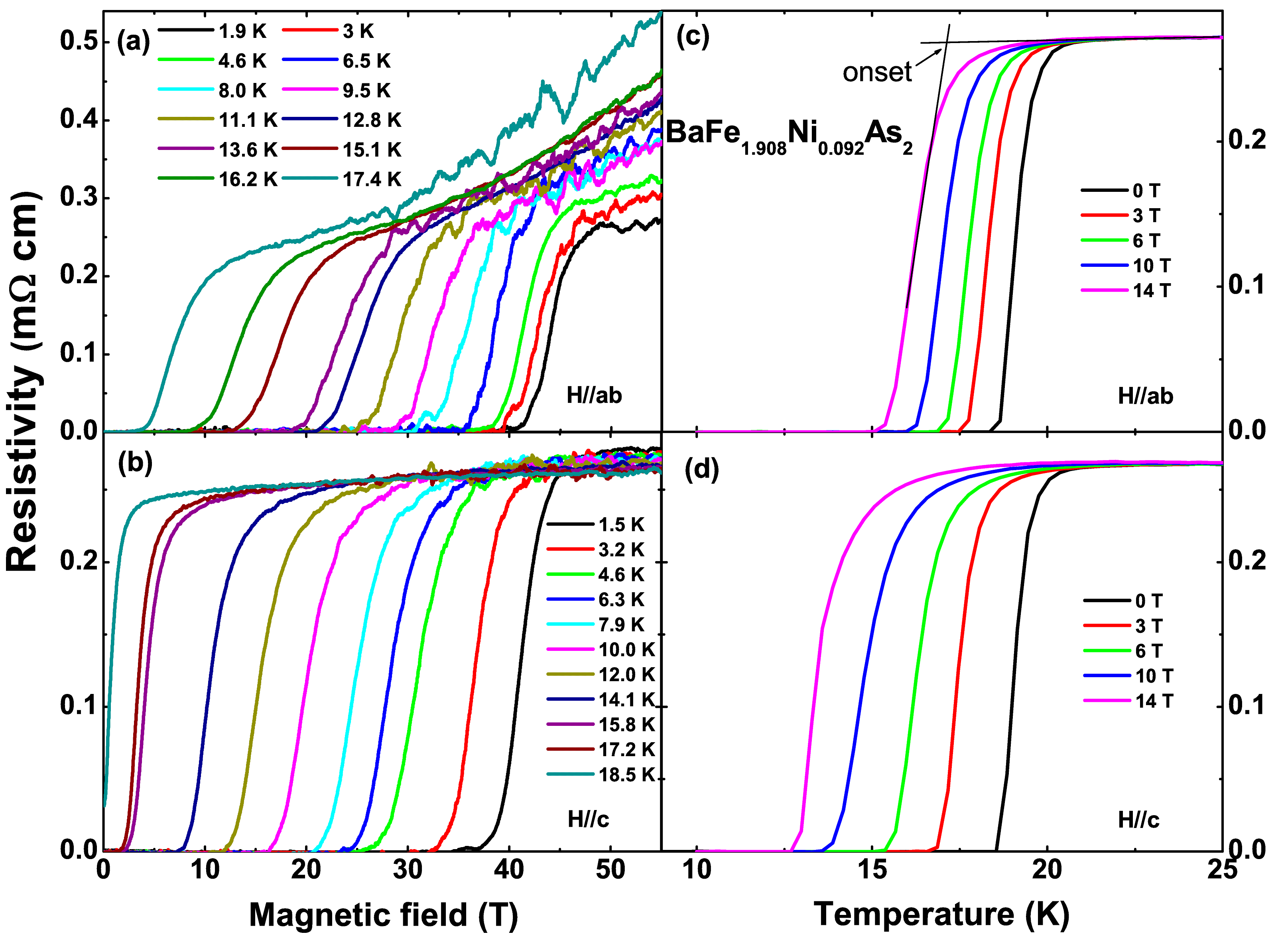}
\caption{(color online) (a)-(b) Magnetic-field and (c)-(d)
temperature dependence  of the in-plane resistivity $\rho_{ab}$ with
$H \parallel ab$ and $H
\parallel c$ of BaFe$_{1.908}$Ni$_{0.092}$As$_2$ measured under pulsed magnetic field and static field in PPMS, respectively.
The upper critical field $H_{c2}$ is defined as the onset of the
superconducting transitions shown in panel (c).} \label{fig:fig2}
\end{figure}

\begin{figure}
\includegraphics[scale=0.55]{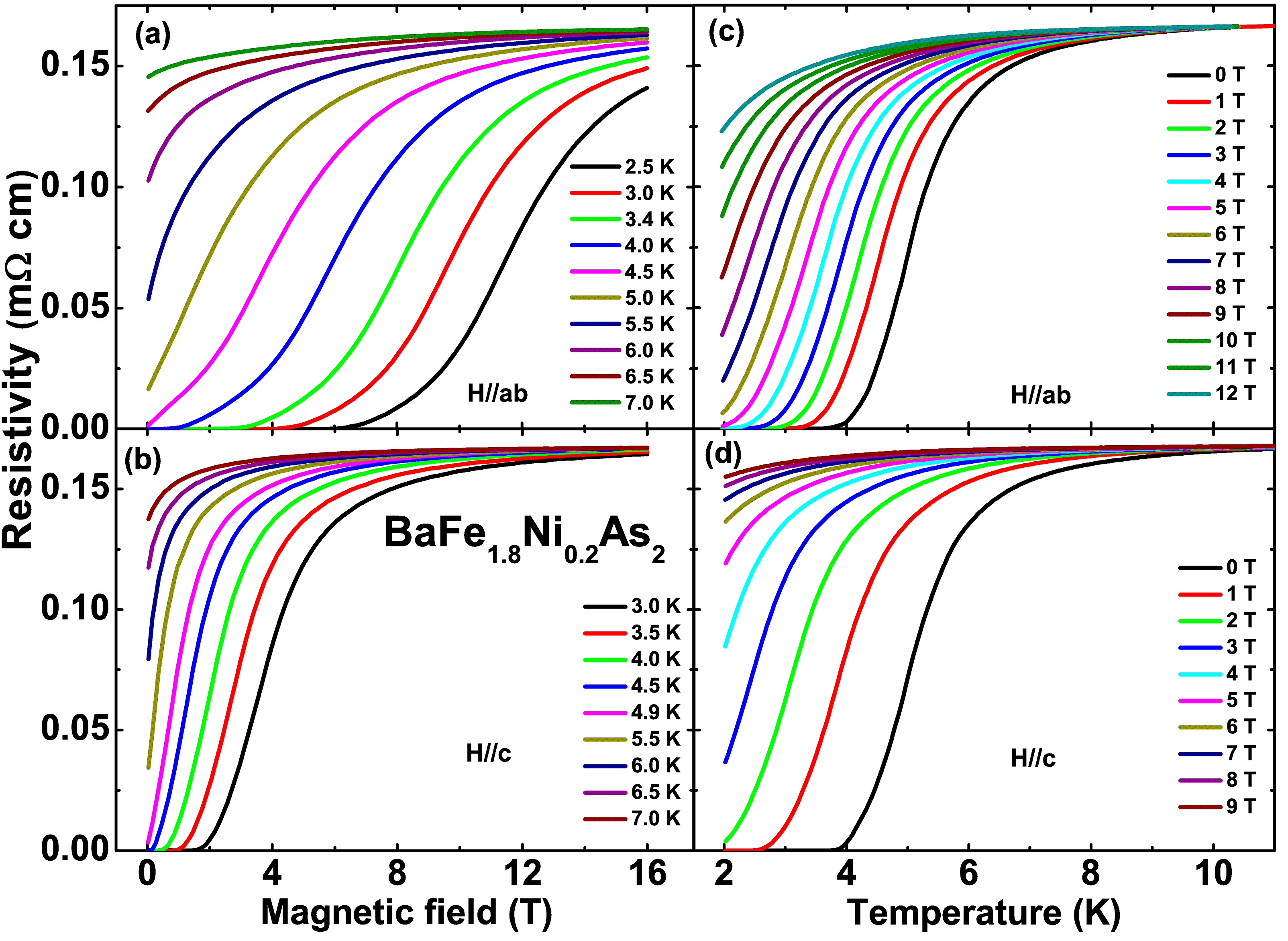}
\caption{(color online) (a) - (d) Magnetic-field and temperature
dependence of $\rho_{ab}$ with $H \parallel ab$ and $H \parallel c$
in BaFe$_{1.80}$Ni$_{0.20}$As$_2$ measured in a 16 T-PPMS.}
\label{fig:fig3}
\end{figure}

Electron doped BaFe$_{2-x}$Ni$_x$As$_2$ single
crystals were grown by the FeAs self-flux method \cite{LuoHQ2}.  Twelve
different compositions across the superconducting dome were
investigated with nominal Ni contents of $x$ = 0.065, 0.075, 0.085, 0.092, 0.096, 0.1, 0.108,
0.12, 0.15, 0.18, 0.20, and 0.22, as marked in Fig. 1(a). Note that the
real Ni concentration is about 0.8 times the nominal content $x$
\cite{LuoHQ2}. All samples show a very narrow superconducting transition width $\Delta T_c\equiv T_c(90\%)-T_c(10\%) < 1$ K
[inset of Fig. 1(a)], indicating a high crystal quality. Detailed characterization of these crystals can be found in our previous publications \cite{LuoHQ1,LuoHQ2}.

The resistivity in the $ab$ plane ($\rho_{ab}$) was measured by a standard
four-probe method with magnetic fields applied parallel to the $ab$
plane ($H\parallel ab$) and the $c$ axis  ($H \parallel c$),
respectively. The field-dependent magnetoresistivity $\rho_{ab}$ was measured at different temperatures using a 65 T
non-destructive pulsed magnet driven by a capacitor bank at the
Dresden High Magnetic Field Laboratory, with a pulse duration of about 180 ms \cite{HLD}. The applied current was 1 mA at a frequency of 30-40 kHz. The voltage was recorded by a digital oscilloscope, Yokogawa DL750, with a high sampling rate of 1 MS/ s and a resolution of 16 bit.  After the pulse, the signal processing is performed by use of a lock-in software procedure. The down-sweep branch of
the pulse was used to determine $H_{c2}$ utilizing its long decay time
(about 150 ms). In order to determine the field dependence of
$H_{c2}$ near $T_c$ more accurately, the temperature dependence of $\rho_{ab}$ was measured
by use of a \emph{Quantum Design} Physical Property Measurement System with
magnetic fields up to 14 T (14 T-PPMS).   Additional data on the $x=0.20$ and 0.22
compounds were measured in a 16 T-PPMS down to 2 K and in pulsed magnetic fields in a He-3 bath cryostat
down to 0.6 K. To ensure a low noise during the measurements, all Ohmic contacts, made by silver epoxy, had a low resistance of less than 1 $ \Omega$.

\begin{figure}
\includegraphics[scale=0.25]{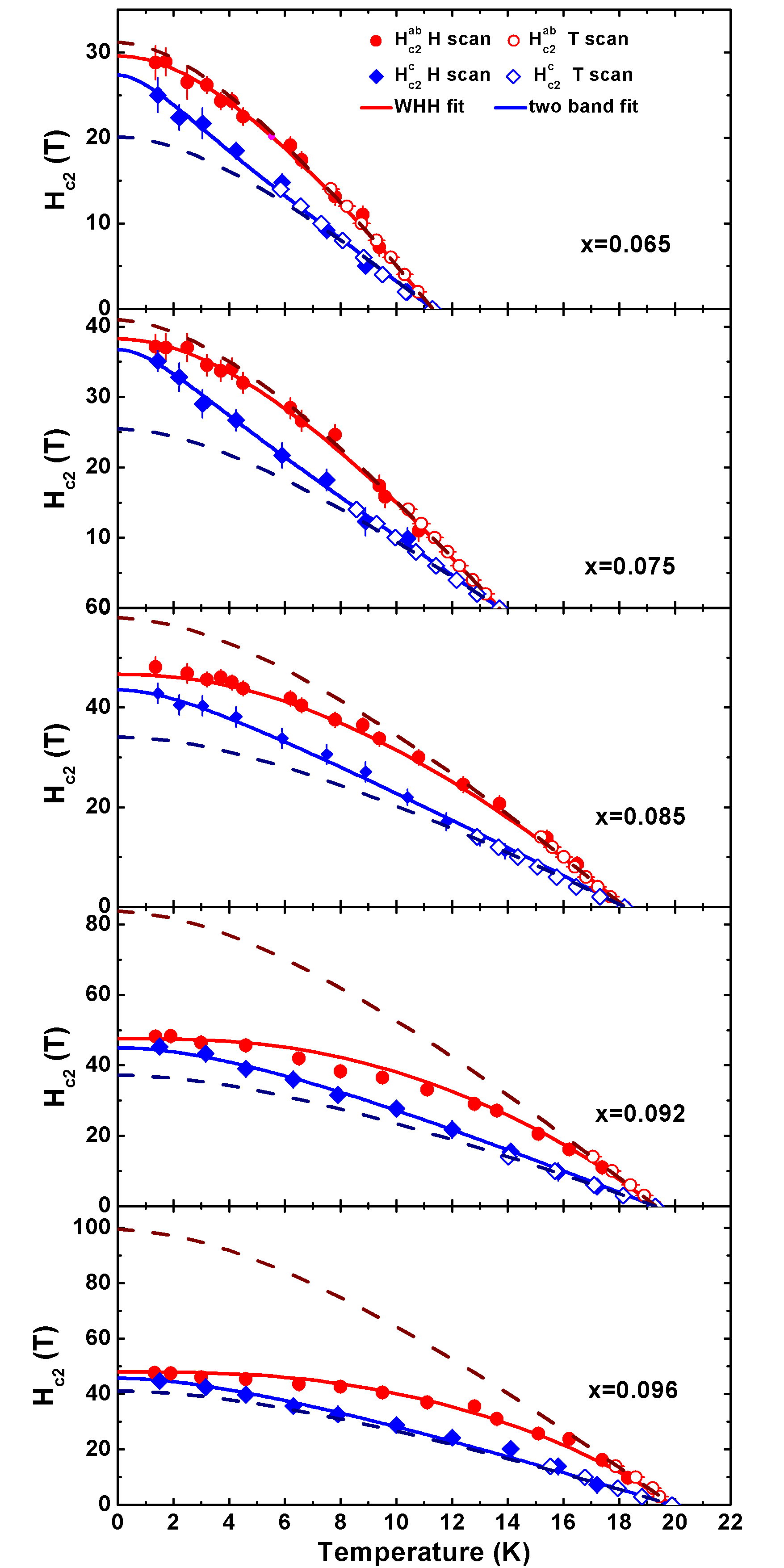}
\caption{(color online) Temperature dependence of $H_{c2}$ of the
five underdoped samples extracted from the magnetotransport
measurements.  The solid symbols are obtained from pulse-field
measurements by scanning field (H scan), and the open symbols are
obtained from PPMS measurements by scanning temperature (T scan).
The red solid line shows a WHH fit with the parameters $\alpha$ and
$\lambda_{so}$ given in Table I for $H_{c2}^{ab}$. The blue solid
line is a two-band fit for $H_{c2}^{c}$ with the parameters $D_1$
and $\eta$ given in Table I. The dashed lines are the WHH
predictions with $\alpha$ = 0 and $\lambda_{so}$ = 0 both for
$H_{c2}^{ab}$ and $H_{c2}^{c}$. } \label{fig:fig4}
\end{figure}

\section{Results and analysis}
We first present typical raw magnetotransport data for the underdoped sample BaFe$_{1.908}$Ni$_{0.092}$As$_2$ with $T_c = 19.3$ K and the overdoped sample BaFe$_{1.8}$Ni$_{0.2}$As$_2$ with $T_c = 6.4$ K in pulsed and static magnetic fields, as shown in Figs. 2 and 3, respectively. For other dopings, the magnetoresistance data are very similar. Thus, we only show the extracted $H_{c2}$ data in the following.

In Fig. 2, only negligible field-induced broadening of the resistive transitions is observed, in contrast to what has
been observed for NdFeAsO$_{0.7}$F$_{0.3}$\cite{Jaroszynski}, SmFeAsO$_{0.85}$, and
SmFeAsO$_{0.8}$F$_{0.2}$\cite{LeeHS}, suggesting a very narrow
vortex-liquid region in the Ba-122 system.  Thus, we determine $H_{c2}$ as
the onset of the transition, most closely corresponding to the
resistive upper critical field \cite{VedeneevSI} [see arrow in Fig. 2(c)].  The full recovery
of the normal-state resistivity allows us to determine $H_{c2}$
quantitatively for both field geometries.
Apparently, a stronger in-plane field ($H \parallel ab$) is needed to suppress
superconductivity, consistent with previous results for iron-based
superconductors\cite{Jaroszynski,LeeHS,HunteF, Altarawneh, YuanHQ,
KanoM, Khim, Khim2,YuanHQ2}. By carefully comparing the results in Fig. 2 and Fig. 3, one may immediately find that the
upper critical field for $H\parallel ab$ ($H_{c2}^{ab}$) is close to the $H\parallel c$ case ($H_{c2}^{c}$) in the underdoped, $x= 0.092$, sample at low temperatures,
while a clear difference between $H_{c2}^{ab}$ and $H_{c2}^{c}$ exists for the overdoped, $x= 0.20$, sample. Therefore, the anisotropy $\gamma=H_{c2}^{ab}/H_{c2}^{c}$, is electron-doping dependent.

The temperature dependence of
$H_{c2}^{ab}$ and $H_{c2}^{c}$ of the five underdoped samples with $x=$ 0.065, 0.075, 0.085, 0.092, and 0.096, is
shown in Fig. 4.   The solid symbols are obtained
from pulsed-field measurements utilizing magnetic field scans (H scan), and the open symbols are obtained from
PPMS measurements by use of temperature scans (T scan).  The consistency of the data from two different measurements proves the reliability of the results.  For all five samples, $H_{c2}^{ab}(T)$ has a tendency to saturate with decreasing temperature, while
$H_{c2}^{c}(T)$ shows a quasilinear increase and no clear saturation
at low temperatures. No obvious upturn of $H_{c2}(T)$ is found for
both field directions, with a nearly isotropic $H_{c2}$ at
1.5 K.  As the doping level becomes higher, $H_{c2}$ at low temperatures becomes more isotropic.

\begin{table*}
\caption{\label{tab:table1}Summary of the parameters for the upper
critical field for all investigated compositions of
BaFe$_{2-x}$Ni$_{x}$As$_2$. }
\begin{ruledtabular}
\begin{tabular}{c c c c c c c c c c c c c c c c}
$x$  &$T_c$  &$-dH_{c2}^{ab}/dT|_{T_{c}}$  &$-dH_{c2}^{c}/dT|_{T_{c}}$ &$H_{c2}^{orb,ab}(0)$ &$H_{c2}^{orb,c}(0)$ &$H_{P}^{BCS}(0)$ &$\alpha$ &$\lambda_{so}$ &$D_1$  &$\eta$ &$H_{c2}^{ab}(0)$  &$H_{c2}^{c}(0)$ &$\gamma(0)$\\
       &(K)        &(T/K)        &(T/K)          &(T) &(T)  &(T)             &                   &            & &$(D_2/D_1)$      &(T)               &(T)\\
\hline
0.065    &11.3       &3.98        &2.57     &31.0  &20.0    &20.8                  &0.33            &0         &4.10      &0.27       &29.6                &27.3        &1.08\\
0.075    &13.7       &4.31        &2.68     &40.7  &25.3    &25.2                  &0.38            &0         &3.70      &0.27       &38.3                &36.7        &1.04\\
0.085    &18.2       &4.59        &2.70     &57.6  &33.9    &33.5                  &0.70            &0.05      &3.45      &0.38        &46.6                &43.5        &1.07\\
0.092    &19.3       &6.25        &2.78     &83.2  &37.0    &35.5                  &1.69            &0.12      &3.15      &0.48        &47.6                &44.9        &1.06\\
0.096    &19.9       &7.20        &2.98     &98.9  &40.9    &36.6                  &2.58            &0.18      &3.10      &0.53        &48.0                &45.7        &1.05\\
0.10     &20.3       &7.69        &3.17     &107.7 &44.4    &38.1                  &2.77            &0.28      &3.00      &0.55        &48.2                &47.5        &1.02\\
0.108    &20.3       &8.42        &3.23     &117.9 &45.2    &38.1                  &3.09            &0.33      &2.95      &0.57        &50.0                &47.6        &1.05\\
0.12     &18.6       &7.77        &2.98     &99.7  &38.2    &34.2                  &2.62            &0.40      &2.88      &0.65        &49.4                &42.6        &1.16\\
0.15     &14.5       &5.20        &2.24     &52.7  &22.7    &26.7                  &1.20            &0.06      &2.88      &2.00        &35.1                &25.6        &1.37\\
0.18     &10.9       &4.36        &2.04     &32.8  &15.3    &20.1                  &0.68            &0.04      &2.92      &3.20        &27.6                &17.8        &1.55\\
0.20     &6.4        &4.35        &1.50     &19.2  &6.5     &11.8                  &0               &0         &3.15      &5.50        &19.3                &9.2         &2.09 \\
0.22     &3.4        &4.25        &1.11     &10.0  &2.6     &6.3                   &0               &0         &3.90      &3.50       &10.1                &4.2         &2.43\\

\end{tabular}
\end{ruledtabular}
\end{table*}

\begin{figure}
\includegraphics[scale=0.25]{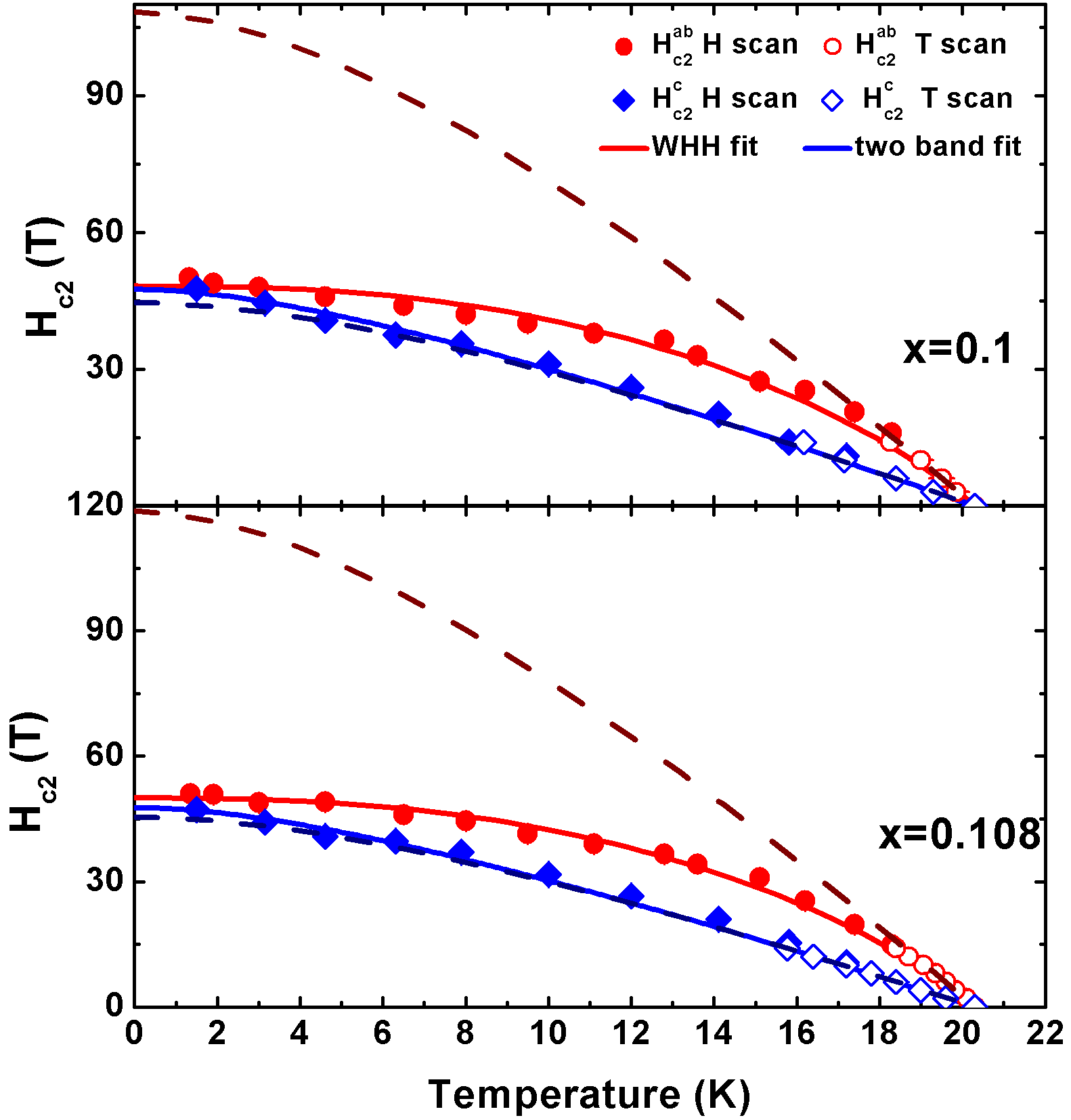}
\caption{(color online) Temperature dependence of $H_{c2}$ of the
two samples at optimum doping extracted from the magnetotransport
measurements.  The solid symbols are obtained from pulse-field
measurements, and the open symbols are obtained from PPMS
measurements.  The red solid line shows a WHH fit with the parameter
$\alpha$ and $\lambda_{so}$ as given in Table I for $H_{c2}^{ab}$.
The blue solid line is a two-band fit for $H_{c2}^{c}$ with the
parameter $D_1$ and $\eta$ given in Table I. The dashed lines are
the WHH predictions with $\alpha$ = 0 and $\lambda_{so}$ = 0 both
for $H_{c2}^{ab}$ and $H_{c2}^{c}$.} \label{fig:fig5}
\end{figure}

In general, by applying a magnetic field on a type-II
superconductor, the Cooper pairs break up via two independent
mechanisms: either by orbital  or by spin-paramagnetic effects. The former is
associated with screening currents around vortex cores in order to expel the external field and to
reduce the condensation energy, while the latter originates from the Zeeman effect. For a single-band
superconductor in the dirty limit \cite{HelfandE}, the orbital limit
is given by $H_{c2}^{orb}(0)=-0.69dH_{c2}/dT|_{T=T_{c}}T_{c}$.  Alternatively, the Pauli-limiting
field for a weakly coupled BCS superconductor in the absence of
spin-orbit scattering can be estimated as \cite{Clogston}
$H_{P}^{BCS}(0)/T_{c} = 1.86$ T/K.  For the typical underdoped sample BaFe$_{1.908}$Ni$_{0.092}$As$_2$ (Fig. 2) for example: we obtain
$-dH_{c2}^{ab}/dT|_{T_{c}} = 6.25 $ K/T and $-dH_{c2}^{c}/dT|_{T_{c}} = 2.78 $ K/T, yielding $H_{c2}^{orb, ab}(0)= 83$ T for $H \parallel ab$,
$H_{c2}^{orb,c}(0)= 37$ T for $H \parallel c$, and $H_{P}^{BCS}(0) = 36$ T.
These estimates do not agree with our experimental data. To
fully describe our results, one must take into account both
orbital pair-breaking and spin-paramagnetic effects.  Therefore, we
use the full Werthamer-Helfand-Hohenberg (WHH) formula that
incorporates the spin paramagnetic effect via the Maki parameter
$\alpha$ and the spin-orbit scattering constant $\lambda_{so}$ to
describe the experimental $H_{c2}$ data\cite{WHH}:
\begin{equation}
\ln\frac{1}{t} =\sum_{\nu=-\infty}^{\infty}\left\{\frac{1}{|2\nu+1|}-[|2\nu+1|+\frac{\bar{h}}{t}+\frac{(\alpha\bar{h}/t)^{2}}{|2\nu+1|+(\bar{h}+\lambda_{so})/t}]^{-1}\right\},
\end{equation}
where $t=T/T_{c}$ and
$\bar{h}=(4/\pi^{2})[H_{c2}(T)/|dH_{c2}/dT|_{T_{c}}]$.  As shown by the
red solid line in Fig. 4, the best fit can reproduce the
experimental $H_{c2}^{ab}$ of all the five samples very well.  All of the fit parameters we used are listed in Table I for all samples. For $x = 0.092$, we obtain
$\alpha=1.69$ and $\lambda_{so}=0.12$.   The Maki parameter
$\alpha$, defined as $\alpha=\sqrt{2}H_{c2}^{orb}(0)/H_{P}(0)$, is
comparable to the cases of LiFeAs\cite{Khim2,YuanHQ2,ChoK},
KFe$_2$As$_2$\cite{Terashima}, and
Ba$_{0.6}$K$_{0.4}$Fe$_2$As$_2$\cite{YuanHQ}, indicating a dominant
spin-paramagnetic effect in the upper critical field for $H
\parallel ab$. In Fig. 4, one can see that the WHH fit(dashed lines) underestimates the low-temperature data of $H_{c2}^{c}(T)$, even when
considering the orbital pair breaking only ($\alpha=\lambda_{so}=0$), while a similar fit heavily overestimates
$H_{c2}^{ab}(T)$ especially for $x=$ 0.092 and 0.096. Thus, the single-band model cannot fully describe $H_{c2}^{c}(T)$ in the underdoped regime.

On the other hand, the quasilinear temperature dependence of
$H_{c2}^{c}$, which has been commonly observed in MgB$_{2}$ and other iron
pnictides \cite{Jaroszynski,LeeHS,HunteF}, can be understood by an
effective two-band model \cite{Gurevich},
\begin{equation}
a_0[\ln{t}+U(h)][\ln{t}+U(\eta h)]+a_1[\ln{t}+U(h)]+a_2[\ln{t}+U(\eta h)]=0.
\end{equation}
The coefficients $a_0$, $a_1$, and $a_2$, are determined from the BCS
coupling tensor $\lambda_{mm'}$\cite{Gurevich}. The function
$U(x)=\psi(1/2+x)-\psi(1/2)$, where $\psi$ is the di-gamma function.
Other parameters are defined by $h=H_{c2}D_1/2\phi_0T$
and $\eta=D_2/D_1$, where $\phi_0$ is the flux quantum and $D_n$ is
the electron diffusivity for the $n$th Fermi-surface sheet. Here, we
assume $a_0=1$, $a_1=1.5$, and $a_2=0.5$ with dominant intraband coupling.  Since the line shape mostly depends on the
choice of $D_1$ and $\eta$ rather than the coupling constants
$\lambda_{mm'}$, we only tune $D_1$ and
$\eta$ to fit $H_{c2}^{c}(T)$, where $\eta \neq 1$ means different intraband
scattering on each Fermi sheet. The two-band fits agree very well with the $H_{c2}^{c}(T)$ data (blue solid lines in Fig. 4). We have also tried to fit $H_{c2}^{ab}(T)$ with a two-band fit, but this does not capture
the saturation of $H_{c2}^{ab}(T)$ due to strong paramagnetic effect. Therefore, we used a single-band WHH fit for $H_{c2}^{ab}(T)$ and two-band fit for $H_{c2}^{c}(T)$ separately, as shown by solid lines in Fig. 4. The best fit parameters are listed in Table I.

We have as well analyzed the data for the optimally doped and overdoped samples in the same way. The fit results are shown in Figs. 5 and 6 with the parameters given in Table I. For the samples at optimum doping (Fig. 5), the single-band WHH fit considering the orbital pair breaking only ($\alpha=\lambda_{so}=0$) overestimates $H_{c2}^{ab}(T)$ heavily as for the underdoped sample, and underestimates the low-temperature data of $H_{c2}^{c}(T)$ slightly. Although a two-band model can better describe the data, the two-band effect may be weaker than for the underdoped sample. Interestingly, $H_{c2}(T)$ at low temperatures becomes very isotropic with $\gamma (0)= 1.02$ for $x=$ 0.10, which is the smallest anisotropy among all measured iron-based superconductors until now \cite{Jaroszynski,LeeHS,HunteF, Altarawneh,YuanHQ, KanoM, Khim, Khim2}.

\begin{figure}
\includegraphics[scale=0.25]{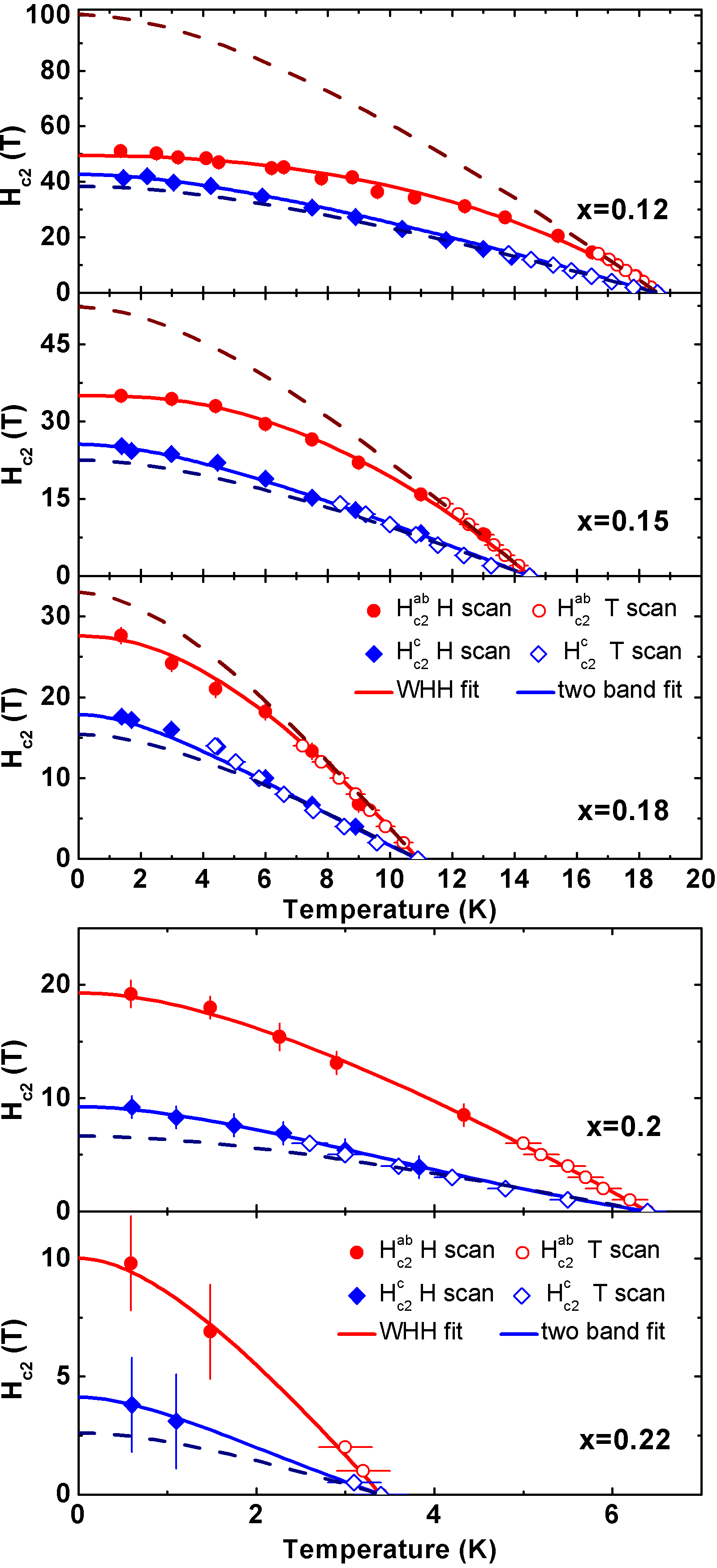}
\caption{(color online) Temperature dependence of $H_{c2}$ of the
five overdoped samples extracted from the magnetotransport
measurements.  The solid symbols are obtained from pulse-field
measurements, and the open symbols are obtained from PPMS
measurements.  The red solid line shows a WHH fit $H_{c2}^{ab}$. The
blue solid line is a two-band fit for $H_{c2}^{c}$. The fit
parameters are given in Table I. The dashed lines are the WHH
predictions with $\alpha$ = 0 and $\lambda_{so}$ = 0 both for
$H_{c2}^{ab}$ and $H_{c2}^{c}$. } \label{fig:fig6}
\end{figure}

The overdoped samples show opposite behavior compared to the underdoped samples.  As shown in Fig. 6,  by increasing Ni doping from $x=$ 0.12 to $x=$ 0.22, the anisotropy of $H_{c2}$ at low temperature grows quickly, and the pure orbital WHH fits (dashed lines) clearly deviate from the $H_{c2}^{c}(T)$ data but agree well with the $H_{c2}^{ab}(T)$ data. Especially for the heavily overdoped sample with $x=$ 0.22, the spin-paramagnetic effects are negligible due to the absent saturation of $H_{c2}^{ab} (T)$. However, we still need the two-band model to describe $H_{c2}^{c} (T)$ in the overdoped regime.  Although the electronic diffusivity
$D_1$ slightly varies between 3 and 4, the diffusivity ratio $\eta=D_2/D_1$ rapidly grows from 0.27 to 5.5 upon
electron doping.  The large increase of the diffusivity ratio $\eta=D_2/D_1$ suggests enhanced electronic
mobility for one of the bands or a significant change of the
relative scattering rate for each band upon electron doping.  By approaching zero temperature, the anisotropy
is beyond 2 for $x=0.20$ and $x=0.22$, clearly different from the
nearly isotropic superconductivity for the underdoped and optimally doped samples.

\section{Discussion}

By fitting the $H_{c2} (T)$ data in Fig. 4 - Fig. 6 (solid lines), we obtain the upper critical field $H_{c2} (0)$ for the zero-temperature limit with $\sim$ 5\% accuracy. We finally summarize the doping dependence of $H_{c2}(0)$ and their anisotropy $\gamma(0)$ in Fig. 7
for the whole superconducting dome from $x = 0.065$ to $ 0.22$.  The gradient color in Fig. 7 maps the temperature dependence of the
anisotropy $\gamma$ for all samples, where all
of them show an increasing anisotropy upon warming to $T_c$. The
overall doping-dependent features of $H_{c2}^{ab}(0)$ and
$H_{c2}^{c}(0)$ follow the superconducting dome, while their
difference quickly increases especially on the overdoped side, resulting
in an abrupt increase of $\gamma(0)$ when $x> 0.1$. As shown in the inset of Fig.
7, the overall doping dependence of $\gamma(0)$ is highly asymmetric with a minimum $\gamma(0)=$ 1.02 at optimal doping $x=$ 0.10,
which should be related to the asymmetric superconducting dome in the BaFe$_{2-x}$Ni$_{x}$As$_2$ system.

It is argued, that the nearly isotropic $H_{c2}$ for
most of the optimally doped iron pnictides may originate either from Pauli-limiting or band-warping effects along the $k_z$ direction\cite{YuanHQ,Khim2,Hosono2,DingH,IdetaS}. In our results, the Pauli-limiting effect (marked as $\alpha$) is indeed strongest around optimum doping, then quickly weakens, and finally vanishes
in the overdoped samples, resulting in small $H_{c2}^{ab}(0)$ and $\gamma(0)$ for optimally doped compounds, and larger
$H_{c2}^{ab}(0)$ and $\gamma(0)$ for the overdoped samples. However,
this simple picture cannot explain the difference in $\gamma$ between the underdoped ($x=0.085$) and the overdoped ($x=0.18$) samples with similar values of $\alpha$ and $\lambda_{so}$ (Table I).
On the other hand, upon doping electrons, the warped cylindrical Fermi surface around the $\Gamma-Z$ line first transforms
to a 3D ellipsoid centered at the Z point, and completely disappears near the concentration where superconductivity appears,
while the electron Fermi-surface sheet around the X point continuously increases\cite{IdetaS,LiuC}. We also notice that the two-band fits of $H_{c2}^{c}(T)$ show a crossover of the diffusivity ratio $\eta=D_2/D_1$ from
$\eta<1$ to $\eta>1$ around $x \approx 0.12$, corresponding to the disappearance of the hole pocket at the $\Gamma$ point. Thus, the Ni dopants remarkably affect the scattering rate and induce an unbalanced mobility between the electron and hole band and additionally change the topology of the Fermi surface. Indeed, recent calculations suggest that the Co/Ni dopants in 122 compounds increase the impurity potential and introduce stronger scattering on the hole band accompanied by a weak interband scattering\cite{KuW,IshidaS,SefatAS}.

Therefore, the doping-dependent anisotropy of the superconductivity in BaFe$_{2-x}$Ni$_{x}$As$_2$ can be understood by the dual effects from the Fermi-surface topology and impurity scattering. At first, the 3D-like hole Fermi surface with dominant diffusivity tends to form a nearly isotropic superconductivity in the underdoped compounds\cite{FZ}. Then the warping effect of the hole sheets becomes more prominent after doping more electrons, and it tends to form an isotropic superconductivity as well. By further doping Ni into the overdoped regime, the charge carriers from the tiny hole pocket around the Z point are mostly localized and contribute insignificantly to the superconductivity. While the less-warped electron Fermi surface with large volume and high mobility gives rise to a deviation from isotropic superconductivity, leading to a highly asymmetric doping dependence of $\gamma(0)$ significantly enhanced in the overdoped regime. Thus, the optimum doping with the most isotropic superconductivity, correlates with an isotropic scattering from different Fermi surfaces with fine-tuned scattering rate and Fermi-surface topology.

\begin{figure}
\includegraphics[scale=0.2]{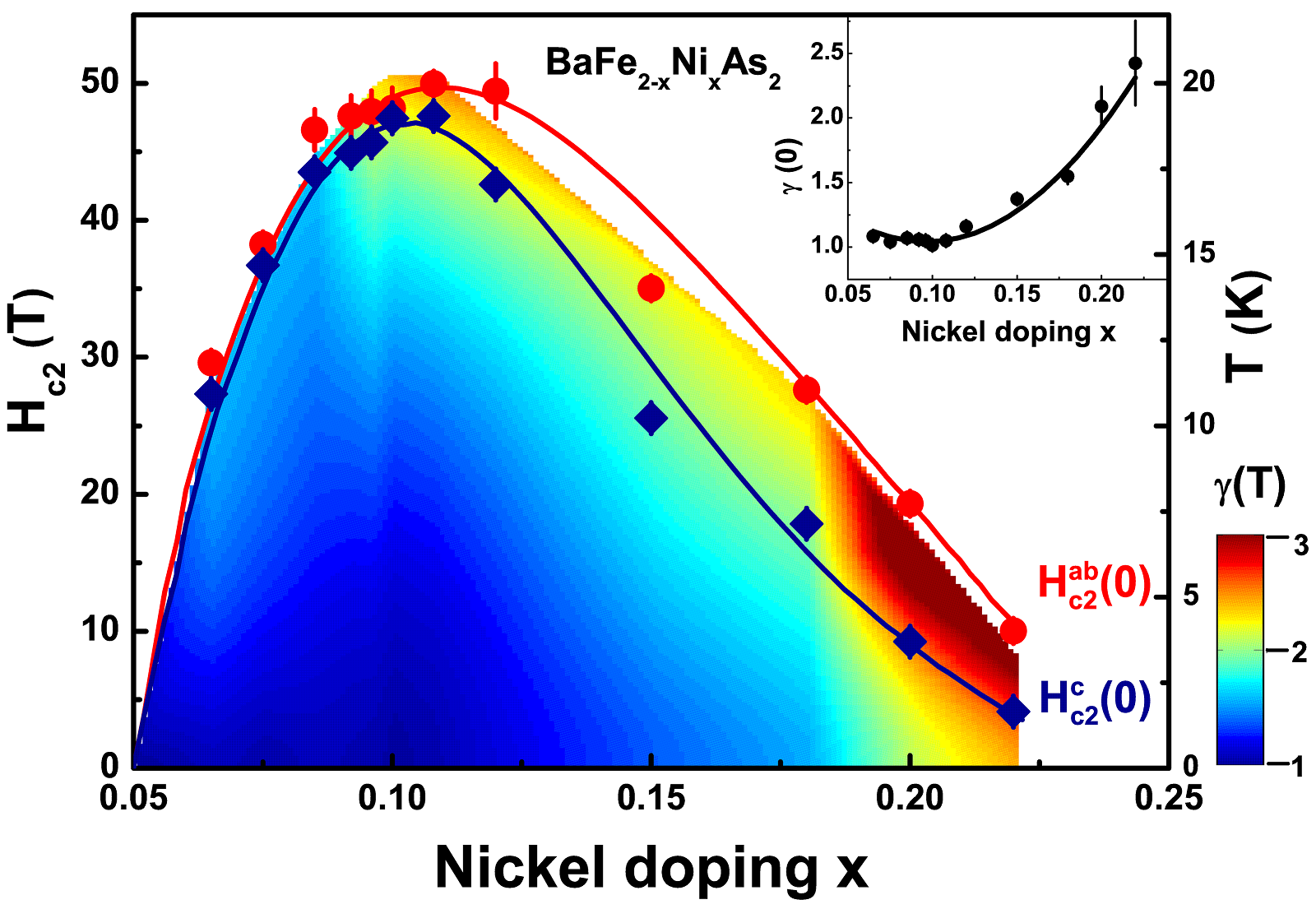}
\caption{(color online) Doping dependence of $H_{c2}^{ab}$,
$H_{c2}^{c}$, and $\gamma$ at 0 K. The gradient color maps the
temperature dependence of the anisotropy $\gamma$ (right axis).}
\label{fig:fig7}
\end{figure}
\section{Summary}

In summary, we have investigated the temperature dependence of the
upper critical field ($H_{c2}$) in a series of
BaFe$_{2-x}$Ni$_{x}$As$_2$ (0.065$\leqslant x \leqslant $0.22) single
crystals in magnetic fields up to 60 T aligned both within the $ab$ plane and along the
$c$ axis.  All $H_{c2}^{ab}(T)$ data can be fitted
by the WHH formula including orbital and spin-paramagnetic
effects, while the quasilinear $H_{c2}^{c}(T)$ data can be
described by an effective two-band model. The anisotropy of
$H_{c2}$ at 0 K, $\gamma(0)$, is close to 1 for the underdoped and
optimally doped samples, but increases beyond 2
for the overdoped samples, forming a highly asymmetric doping dependence similar to the superconducting dome.
Our results indicate that the superconducting anisotropy is determined by the topology of the Fermi surfaces together with the doping-induced impurity
scattering, and that the multi-band physics is important for the optimal superconductivity in iron pnictides.

We acknowledge the support of the HLD at HZDR, member of the European Magnetic Field Laboratory (EMFL).
The work at IOP,  CAS, is supported by MOST (973 project: 2011CBA00110 and 2012CB821400), NSFC (11374011 and 11374346), and CAS (SPRP-B: XDB07020300).

{}


\begin{thebibliography}{}

\bibitem{Hosono1} Y. Kamihara, T. Watanabe, M. Hirano, and H. Hosono, J. Am. Chem. Soc. \textbf{130}, 3296-3297 (2008).

\bibitem{WangZSPRB} Z. S. Wang, H. Q. Luo, C. Ren, and H. H. Wen, Phys. Rev. B \textbf{78}, 140501 (2008).


\bibitem{Jaroszynski} J. Jaroszynski, F. Hunte, L. Balicas, Y. -j. Jo, I. Rai\v{c}evi\v{c}, A. Gurevich, D.C. Larbalestier, F.F. Balakirev, L. Fang, P. Cheng, Y. Jia, and H. -H. Wen, Phys. Rev. B \textbf{78}, 174523(2008).


\bibitem{LeeHS} H. -S. Lee, M. Bartkowiak, J. -H. Park, J. -Y. Lee, J. -Y. Kim, N. -H. Sung, B.K. Cho, C.-U. Jung, J.S. Kim, and H. -J. Lee, Phys. Rev. B \textbf{80}, 144512 (2009).


\bibitem{HunteF} F. Hunte, J. Jaroszynski, A. Gurevich, D. . Larbalestier, R. Jin, A.S. Sefat, M. A. McGuire, B. C. Sales, D. K. Christen, and D. Mandrus, Nature \textbf{453}, 903 (2008).

\bibitem{Altarawneh} M. M. Altarawneh, K. Collar, C. H. Mielke, N. Ni, S. L. Budko, and P. C. Canfield, Phys. Rev. B \textbf{78}, 220505 (2008).

\bibitem{YuanHQ} H. Yuan, J. Singleton, F. F. Balakirev, S. A. Baily, G. Chen, J. Luo, and N. Wang, Nature \textbf{457}, 565 (2009).


\bibitem{KanoM} M. Kano, Y. Kohama, D. Graf, F. Balakirev, A. S. Sefat, M. A. Mcguire, B. C. Sales, D. Mandrus, and S. W. Tozer,, J. Phys. Soc. Jpn. \textbf{78}, 084719 (2009).

\bibitem{Khim} S. Khim, J. W. Kim, E. S. Choi, Y. Bang, M. Nohara, H. Takagi, and K.H. Kim, Phys. Rev. B \textbf{81},184511 (2010).

\bibitem{Khim2} S. Khim, B. Lee, J. W. Kim, E. S. Choi, G. R. Stewart, and K. H. Kim, Phys. Rev. B \textbf{84}, 104502 (2011).

\bibitem{YuanHQ2} J. L. Zhang, L. Jiao, F. F. Balakirev, X. C. Wang, C. Q. Jin, and H. Q. Yuan, Phys. Rev. B \textbf{83}, 174506 (2011).

\bibitem{RotterM1} M. Rotter, M. Tegel, D. Johrendt, I. Schellenberg, W. Hermes, and R. P\"{o}ttgen, Phys. Rev. B \textbf{78}, 020503(R) (2008).

\bibitem{RotterM2} M. Rotter, M. Tegel, and D. Johrendt, Phys. Rev. Lett. \textbf{101}, 107006 (2008).


\bibitem{XuZA} L. J. Li, Y. K. Luo, Q. B. Wang, H. Chen, Z. Ren, Q. Tao, Y. K. Li, X. Lin, M. He, Z. W. Zhu, G. H. Cao, and Z. -A. Xu, New J. Phys. \textbf{11} 025008 (2009).

\bibitem{NiN} N. Ni, A. Thaler, J. Q. Yan, A. Kracher, E. Colombier, S. L. Bud'ko, P. C. Canfield, and S. T. Hannahs, Phys. Rev. B \textbf{82}, 024519 (2010).

\bibitem{MandrusD} D. Mandrus, A. S. Sefat, M. A. McGuire, and B. C. Sales, Chem. Mater. \textbf{22}, 715 (2010).


\bibitem{AswarthamS} S. Aswartham, M. Abdel-Hafiez, D. Bombor, M. Kumar, A. U. B. Wolter, C. Hess, D. V. Evtushinsky, V. B. Zabolotnyy, A. A. Kordyuk, T. K. Kim, S. V. Borisenko, G. Behr, B. B\"{u}chner, and S. Wurmehl, Phys. Rev. B \textbf{85}, 224520 (2012).


\bibitem{LuoHQ1} H. Luo, R. Zhang, M. Laver, Z. Yamani, M. Wang, X. Lu, M. Wang, Y. Chen, S. Li, S. Chang, J. W. Lynn, and P. Dai, Phys. Rev. Lett. \textbf{108}, 247002 (2012).


\bibitem{LuXY} X. Lu, H. Gretarsson, R. Zhang, X. Liu, H. Luo, W. Tian, M. Laver, Z. Yamani, Y. -J. Kim, A. H. Nevidomskyy, Q. Si, and P. Dai, Phys. Rev. Lett. \textbf{110}, 257001 (2013).

\bibitem{DingH} P. Richard, T. Qian, and H. Ding, arXiv: 1503.07269.

\bibitem{KuW} T. Berlijn, C. -H. Lin, W. Garber, and W. Ku, Phys. Rev. Lett. \textbf{108}, 207003 (2012).


\bibitem{BlackburnS} S. Blackburn, B. Pr\'{e}vost, M. Bartkowiak, O. Ignatchik, A. Polyakov, T. F\"{o}rster, M. C\^{o}t\'{e}, G. Seyfarth, C. Capan, Z. Fisk, R. G. Goodrich, I. Sheikin, H. Rosner, A. D. Bianchi, and J. Wosnitza, Phys. Rev. B \textbf{89}, 220505(R) (2014).


\bibitem{IdetaS} S. Ideta, T. Yoshida, I. Nishi, A. Fujimori, Y. Kotani, K. Ono, Y. Nakashima, S. Yamaichi, T. Sasagawa, M. Nakajima, K. Kihou, Y. Tomioka, C. H. Lee, A. Iyo, H. Eisaki, T. Ito, S. Uchida, and R. Arita, Phys. Rev. Lett. \textbf{110}, 107007 (2013).


\bibitem{LiuC} C. Liu, A. D. Palczewski, R. S. Dhaka, T. Kondo, R. M. Fernandes, E. D. Mun, H. Hodovanets, A. N. Thaler, J. Schmalian, S. L. Bud'ko, P. C. Canfield, and A. Kaminski, Phys. Rev. B \textbf{84}, 020509(R) (2011).


\bibitem{LiuY} Y. Liu, M. A. Tanatar, W. E. Straszheim, B. Jensen, K. W. Dennis, R. W. McCallum, V. G. Kogan, R. Prozorov, and T. A. Lograsso, Phys. Rev. B \textbf{89}, 134504 (2014).


\bibitem{ZhangS} S. Zhang, Y. P. Singh, X. Y. Huang, X. J. Chen, M. Dzero, C. C. Almasan, arXiv:1507.03628.

\bibitem{KhanSN} S.N. Khan and D.D. Johnson, Phys. Rev. Lett. \textbf{112}, 156401 (2014).

\bibitem{LuoHQ2} Y. Chen, X. Lu, M. Wang, H. Luo, and S. Li, Supercond. Sci. Technol. \textbf{24}, 065004 (2011).

\bibitem{HLD} S. Zherlitsyn, T. Herrmannsdoerfer, Yu. Skourski, A. Sytcheva, and J. Wosnitza, J. Low Temp. Phys. \textbf{146}, 719 (2007).




\bibitem{VedeneevSI} S. I. Vedeneev, B. A. Piot, D. K. Maude, and A. V. Sadakov, Phys. Rev. B \textbf{87},134512 (2013).


\bibitem{Mosqueira1} R. I. Rey, C. Carballeira, J. Mosqueira, S. Salem-Sugui Jr, A. D. Alvarenga, H. Luo, X. Lu, Y. Chen and F. Vidal, Supercond. Sci. Technol. \textbf{26}, 055004 (2013).

\bibitem{HelfandE} E. Helfand and N. R. Werthamer, Phys. Rev. \textbf{147}, 288 (1966).

\bibitem{Clogston} A. M. Clogston, Phys. Rev. Lett. \textbf{9}, 266 (1962).


\bibitem{WHH} N. Werthamer, E. Helfand, and P. Hohenberg, Phys. Rev. \textbf{147}, 295 (1966).

\bibitem{ChoK} K. Cho, H. Kim, M.A. Tanatar, Y. J. Song, Y. S. Kwon, W.A. Coniglio, C. C. Agosta, A. Gurevich, and R. Prozorov, Phys. Rev. B \textbf{83}, 060502 (2011).


\bibitem{Terashima} T. Terashima, M. Kimata, H. Satsukawa, A. Harada, K. Hazama, S. Uji, H. Harima, G. Chen, J. Luo, and N. Wang, J. Phys. Soc. Jpn.\textbf{78}, 063702 (2009).

\bibitem{Gurevich} A. Gurevich, Phys. Rev. B \textbf{67}, 184515 (2003).

\bibitem{Hosono2} S. A. Baily, Y. Kohama, H. Hiramatsu, B. Maiorov, F. F. Balakirev, M. Hirano, and H. Hosono, Phys. Rev. Lett. \textbf{102}, 117004 (2009).


\bibitem{IshidaS} S. Ishida, M. Nakajima, T. Liang, K. Kihou, C. H. Lee, A. Iyo, H. Eisaki, T. Kakeshita, Y. Tomioka, T. Ito, and S. Uchida, Phys. Rev. Lett. \textbf{110}, 207001 (2013).

\bibitem{SefatAS} A. S. Sefat, R. Jin, M. A. McGuire, B. C. Sales, D. J. Singh, and D. Mandrus, Phys. Rev. Lett. \textbf{101}, 117004 (2008).

\bibitem{FZ} G. T. Wang, Y. Qian, G. Xu, X. Dai, and Z. Fang, Phys. Rev. Lett. \textbf{104}, 047002 (2010).

\end{thebibliography}
\end{document}